Relativistic five-quark equations and the strange (S = - 3) pentaquarks.

S.M. Gerasyuta and V.I. Kochkin.

Department of Physics, St. Petersburg State Forest Technical University, Institutski Per. 5, St. Petersburg 194021, Russia, e-mail: serg.gerasyuta@gmail.com.

Abstract

The relativistic five-quark equations are found in the framework of the dispersion relation technique. The solutions of these equations using the method based on the extraction of the leading singularities of the amplitudes are obtained. The five-quark amplitudes for the $S = -3$ strange pentaquarks including the $u$, $d$, $s$ quarks determine the masses of strange pentaquarks. The mass spectra of the strange pentaquarks with $J^P = \frac{1^{\pm}}{2}, \frac{3^{\pm}}{2}$ are calculated.

The exotic hadrons observed in experiments are important. They are expected the properties such as quantum numbers, masses, and decay patterns, cannot the explained by the ordinary picture of hadrons, baryons as $qqq$ and mesons as $q\bar{q}$.

The discovery of $P_c^+$, $P_c^{+*}$ [1] has triggered extensive theoretical studies to understand the structure of these pentaquarks [2 – 12]. They are kinematic effects due to rescattering among different channels [2 – 5]. On the other hand, these pentaquarks could be bound states formed from open-charm baryon and charmed meson [6 – 10]. The exotic baryons with hidden charm as antiquark-diquark-diquark are obtained [11, 12].

In our papers [13 – 16], relativistic generalization of three-body Faddeev equations was obtained in the form of dispersion relations in the pair energy of two interacting quarks. We searched for the approximate solution of integral three-quark equations by taking into account two-particle and triangle singularities, all the weaker ones being neglected.

In the present paper, the relativistic five-quark equations are found in the framework of couple-channel formalism. In this work we focus on the strange ($S = -3$) pentaquarks states $sssu\bar{d}$ with the isospin $I = 0$ and spin-parity $J^P = \frac{1}{2}^{\pm}, \frac{3}{2}^{\pm}$. We calculated the masses of pentaquarks $sssu\bar{d}$ with the strangeness $S = -3$ [Table I]. Experimental masses of strange pentaquarks are given in parentheses [17].

The relativistic five-quark equations in the framework of the dispersion relation technique are derived. We use only planar diagrams; the other diagrams due to the rules of $1/N_c$ expansion [18 – 20] are neglected. The current generates a five-quark system. Their successive pair interactions lead to the diagrams shown in Fig. 1. The correct equations for the amplitude are obtained by taking into account all possible subamplitudes. Then one should represent a five-particle amplitude as a sum of ten subamplitudes:

$$A = A_{12} + A_{13} + A_{14} + A_{15} + A_{23} + A_{24} + A_{25} + A_{34} + A_{35} + A_{45}.$$

This defines the division of the diagrams into a group according to the certain pair interaction of particles. The total amplitude can be represented graphically as a sum of diagrams.

We need to consider only one group of diagrams and the amplitude corresponding to them, for example, $A_{12}$. For the sake of simplicity we shall consider the derivation of the relativistic generalization of the Faddeev-Yakubovsky approach for the example of pentaquark. The set of diagrams associated with the amplitude $A_{12}$ can further be broken down into groups corresponding to amplitudes: $A_1(s, s_{1234}, s_{13}, s_{25})$, $A_2(s, s_{1234}, s_{12}, s_{34})$, $A_3(s, s_{1234}, s_{25}, s_{34})$, $A_4(s, s_{1234}, s_{13}, s_{123})$, $A_5(s, s_{1234}, s_{23}, s_{123})$, $A_6(s, s_{1234}, s_{35}, s_{135})$ (Fig. 1). The coefficients are determined by the permutation of quarks [21, 22].

In order to represent the subamplitudes $A_1(s, s_{1234}, s_{13}, s_{25})$, $A_2(s, s_{1234}, s_{12}, s_{34})$, $A_3(s, s_{1234}, s_{25}, s_{34})$, $A_4(s, s_{1234}, s_{13}, s_{123})$, $A_5(s, s_{1234}, s_{23}, s_{123})$ and $A_6(s, s_{1234}, s_{35}, s_{135})$ in the form of a dispersion relation, it is necessary to define the amplitudes of quark-quark and quark-antiquark interaction $b_n(s_{ik})$. The pair quarks amplitudes $q\bar{q} \to q\bar{q}$ and $qq \to qq$ are calculated in the framework of the dispersion N/D method with the input four-fermion interaction with quantum numbers of the gluon [23]. We use the results of our relativistic quark model and write down the pair quarks amplitude in the form:

$$b_n(s_{ik}) = \frac{G_n^2(s_{ik})}{1 - B_n(s_{ik})}, \qquad (1)$$

$$B_n(s_{ik}) = \int_{(m_1+m_2)^2}^{\Lambda_n} \frac{ds'_{ik}}{\pi} \frac{\rho_n(s'_{ik}) G_n^2(s'_{ik})}{s'_{ik} - s_{ik}}. \qquad (2)$$

Here $s_{ik}$ is the two-particle subenergy squared, $s_{ijk}$ corresponds to the energy squared of particles $i$, $j$, $k$, $s_{ijkl}$ is the four-particle subenergy squared and $s$ is the system total energy squared. $G_n(s_{ik})$ are the quark-quark and quark-antiquark vertex functions. $B_n(s_{ik})$ is the Chew-Mandelstam function with cut-off $\Lambda_n$ ($\Lambda_1 = \Lambda_3$) [15].

The n=1 corresponds to a $qq$-pair with $J^P = 0^+$ in the $\bar{3}_c$ color state, n=2 describes a $qq$-pair with $J^P = 1^+$ in the $\bar{3}_c$ color state and n=3 defines the $q\bar{q}$-pairs corresponding to mesons with quantum numbers: $J^{PC} = 0^{++}, 0^{-+}$.

In the case in question the interacting quarks do not produce a bound state, therefore the integration in Eqs. (3) - (8) is carried out from the threshold $(m_i + m_k)^2$ to the cut-off $\Lambda_n$. The system of integral equations, corresponding to Fig. 1 (the meson state with $J^{PC} = 0^{++}$ and diquarks with $J^P = 0^+, 1^+$) can be described as:

$$A_1(s, s_{1234}, s_{13}, s_{25}) = \frac{\lambda_1 B_3(s_{13}) B_1(s_{25})}{[1 - B_3(s_{13})][1 - B_1(s_{25})]} + 3\hat{J}_2(3,1) A_5(s, s_{1234}, s'_{23}, s'_{234}) +$$
$$+ 3\hat{J}_2(3,1) A_6(s, s_{1234}, s'_{35}, s'_{345}) + 2\hat{J}_2(3,1) A_4(s, s_{1234}, s'_{12}, s'_{123}) +$$
$$+ 2\hat{J}_1(3) A_1(s, s_{1234}, s'_{13}, s_{134}) + 2\hat{J}_1(2) A_3(s, s_{1234}, s'_{13}, s_{134}) + \qquad (3)$$
$$+ 2\hat{J}_1(1) A_1(s, s_{1234}, s'_{25}, s_{245}) + 2\hat{J}_1(2) A_2(s, s_{1234}, s'_{25}, s_{245})$$

$$A_2(s, s_{1234}, s_{12}, s_{34}) = \frac{\lambda_2 B_3(s_{12}) B_2(s_{34})}{[1 - B_3(s_{12})][1 - B_2(s_{34})]} + 6\hat{J}_2(3,2) A_5(s, s_{1234}, s'_{23}, s'_{234}) +$$
$$+ 2\hat{J}_2(3,2) A_4(s, s_{1234}, s'_{13}, s'_{134}) + \qquad (4)$$
$$+ 2\hat{J}_1(3) A_2(s, s_{1234}, s'_{12}, s_{125}) + 2\hat{J}_1(1) A_3(s, s_{1234}, s'_{12}, s_{125}) +$$
$$+ 4\hat{J}_1(2) A_2(s, s_{1234}, s'_{34}, s_{345})$$

$$A_3(s, s_{1234}, s_{25}, s_{34}) = \frac{\lambda_3 B_1(s_{25}) B_2(s_{34})}{[1 - B_1(s_{25})][1 - B_2(s_{34})]} + 6\hat{J}_2(1,2) A_5(s, s_{1234}, s'_{25}, s'_{235}) +$$
$$+ 6\hat{J}_2(1,2) A_6(s, s_{1234}, s'_{35}, s'_{235}) + \qquad (5)$$
$$+ 4\hat{J}_1(3) A_1(s, s_{1234}, s'_{34}, s_{134}) + 4\hat{J}_1(3) A_2(s, s_{1234}, s'_{25}, s_{125})$$

$$A_4(s, s_{1234}, s_{13}, s_{123}) = \frac{\lambda_4 B_3(s_{13})}{1 - B_3(s_{13})} + 8\hat{J}_3(3) A_2(s, s_{1234}, s'_{12}, s'_{34}) + 4\hat{J}_3(3) A_1(s, s_{1234}, s'_{23}, s'_{14}), \quad (6)$$

$$A_5(s, s_{1234}, s_{23}, s_{123}) = \frac{\lambda_5 B_1(s_{23})}{1 - B_1(s_{23})} + 4\hat{J}_3(1) A_3(s, s_{1234}, s'_{25}, s'_{34}) +$$
$$+ 2\hat{J}_3(1) A_2(s, s_{1234}, s'_{12}, s'_{34}) + 2\hat{J}_3(1) A_1(s, s_{1234}, s'_{24}, s'_{13})$$
(7)

$$A_6(s, s_{1234}, s_{35}, s_{135}) = \frac{\lambda_6 B_2(s_{35})}{1 - B_2(s_{35})} + 4\hat{J}_3(2) A_3(s, s_{1234}, s'_{34}, s'_{25}) +$$
$$+ 2\hat{J}_3(2) A_1(s, s_{1234}, s'_{23}, s'_{15}) + 2\hat{J}_3(2) A_2(s, s_{1234}, s'_{34}, s'_{15})$$
(8)

where $\lambda_i$ are the current constants. We introduce the integral operators:

$$\hat{J}_1(l) = \frac{G_l(s_{12})}{[1 - B_l(s_{12})]} \int_{(m_1+m_2)^2}^{\Lambda_l} \frac{ds'_{12}}{\pi} \frac{G_l(s'_{12}) \rho_l(s'_{12})}{s'_{12} - s_{12}} \int_{-1}^{+1} \frac{dz_1}{2},$$
(9)

$$\hat{J}_2(l, p) = \frac{G_l(s_{12}) G_p(s_{34})}{[1 - B_l(s_{12})][1 - B_p(s_{34})]} \times$$
$$\times \int_{(m_1+m_2)^2}^{\Lambda_l} \frac{ds'_{12}}{\pi} \frac{G_l(s'_{12}) \rho_l(s'_{12})}{s'_{12} - s_{12}} \int_{(m_3+m_4)^2}^{\Lambda_p} \frac{ds'_{34}}{\pi} \frac{G_p(s'_{34}) \rho_p(s'_{34})}{s'_{34} - s_{34}} \int_{-1}^{+1} \frac{dz_3}{2} \int_{-1}^{+1} \frac{dz_4}{2},$$
(10)

$$\hat{J}_3(l) = \frac{G_l(s_{12}, \tilde{\Lambda})}{1 - B_l(s_{12}, \tilde{\Lambda})} \times$$
$$\times \frac{1}{4\pi} \int_{(m_1+m_2)^2}^{\tilde{\Lambda}} \frac{ds'_{12}}{\pi} \frac{G_l(s'_{12}, \tilde{\Lambda}) \rho_l(s'_{12})}{s'_{12} - s_{12}} \int_{-1}^{+1} \frac{dz_1}{2} \int_{-1}^{+1} dz \int_{z_2^-}^{z_2^+} dz_2 \frac{1}{\sqrt{1 - z^2 - z_1^2 - z_2^2 + 2zz_1z_2}},$$
(11)

where $l, p$ are equal 1 - 3. $m_i$ is a quark mass.

In Eqs. (9) and (11) $z_1$ is the cosine of the angle between the relative momentum of the particles 1 and 2 in the intermediate state and the momentum of the particle 3 in the final state, taken in the c.m. of particles 1 and 2. In Eq. (11) $z$ is the cosine of the angle between the momenta of the particles 3 and 4 in the final state, taken in the c.m. of particles 1 and 2. $z_2$ is the cosine of the angle between the relative momentum of particles 1 and 2 in the intermediate state and the momentum of the particle 4 in the final state, is taken in the c.m. of particles 1 and 2. In Eq. (10): $z_3$ is the cosine of the angle between relative momentum of particles 1 and 2 in the intermediate state and the relative momentum of particles 3 and 4 in the intermediate state, taken in the c.m. of particles 1 and 2. $z_4$ is the cosine of the angle between the relative momentum of the particles 3 and 4 in the intermediate state and that of the momentum of the particle 1 in the intermediate state, taken in the c.m. of particles 3, 4.

We can pass from the integration over the cosines of the angles to the integration over the subenergies [24].

Let us extract two-particle singularities in the amplitudes $A_1(s,s_{1234},s_{13},s_{25})$, $A_2(s,s_{1234},s_{12},s_{34})$, $A_3(s,s_{1234},s_{25},s_{34})$, $A_4(s,s_{1234},s_{13},s_{123})$, $A_5(s,s_{1234},s_{23},s_{123})$ and $A_6(s,s_{1234},s_{35},s_{135})$:

$$A_1(s,s_{1234},s_{13},s_{25}) = \frac{\alpha_1(s,s_{1234},s_{13},s_{25})B_3(s_{13})B_1(s_{25})}{[1-B_3(s_{13})][1-B_1(s_{25})]}, \tag{12}$$

$$A_2(s,s_{1234},s_{12},s_{34}) = \frac{\alpha_2(s,s_{1234},s_{12},s_{34})B_3(s_{12})B_2(s_{34})}{[1-B_3(s_{12})][1-B_2(s_{34})]}, \tag{13}$$

$$A_3(s,s_{1234},s_{25},s_{34}) = \frac{\alpha_3(s,s_{1234},s_{25},s_{34})B_1(s_{25})B_2(s_{34})}{[1-B_1(s_{25})][1-B_2(s_{34})]} \tag{14}$$

$$A_4(s,s_{1234},s_{13},s_{123}) = \frac{\alpha_4(s,s_{1234},s_{13},s_{123})B_3(s_{13})}{1-B_3(s_{13})}, \tag{15}$$

$$A_5(s,s_{1234},s_{23},s_{123}) = \frac{\alpha_5(s,s_{1234},s_{23},s_{123})B_1(s_{23})}{1-B_1(s_{23})}. \tag{16}$$

$$A_6(s,s_{1234},s_{35},s_{135}) = \frac{\alpha_6(s,s_{1234},s_{35},s_{135})B_2(s_{35})}{1-B_2(s_{35})} \tag{17}$$

We do not extract three- and four-particle singularities, because they are weaker than two-particle singularities.

We used the classification of singularities, which was proposed in paper [24]. The construction of approximate solution of Eqs. (3) - (8) is based on the extraction of the leading singularities of the amplitudes. The main singularities in $s_{ik} \approx (m_i + m_k)^2$ are from pair rescattering of the particles i and k. First of all, there are threshold square-root singularities. Also possible are pole singularities, which correspond to the bound states. The diagrams of Fig.1 apart from two-particle singularities have the triangular singularities, the singularities defining the interaction of four and five particles. Such classification allows us to search the corresponding solution of Eqs. (3) - (8) by taking into account some definite number of leading singularities and neglecting all the weaker ones. We consider the approximation, which defines two-particle, triangle, four- and five-particle singularities. The functions $\alpha_1(s,s_{1234},s_{13},s_{25})$, $\alpha_2(s, s_{1234}, s_{12}, s_{34})$, $\quad \alpha_3(s, s_{1234}, s_{25}, s_{34})$, $\quad \alpha_4(s, s_{1234}, s_{13}, s_{123})$, $\quad \alpha_5(s, s_{1234}, s_{23}, s_{123})$ and $\alpha_6(s, s_{1234}, s_{35}, s_{135})$ are smooth functions of $s_{ik}$, $s_{ijk}$, $s_{ijkl}$, $s$ as compared with the singular part of the amplitudes, hence they can be expanded in a series in the singularity point and only the first term of this series should be employed further. Using this classification one define the reduced amplitudes $\alpha_1$, $\alpha_2$, $\alpha_3$, $\alpha_4$, $\alpha_5$, $\alpha_6$ as well as the B functions in the middle point of the physical region of Dalitz-plot at the point $s_0$:

$$s_0^{ik} = s_0 = \frac{s + 3\sum_{i=1}^{5} m_i^2}{0.25 \sum_{\substack{i,k=1 \\ i \neq k}}^{5} (m_i + m_k)^2} \tag{18}$$

$$s_{123} = 0.25 s_0 \sum_{\substack{i,k=1 \\ i \neq k}}^{3} (m_i + m_k)^2 - \sum_{i=1}^{3} m_i^2, \quad s_{1234} = 0.25 s_0 \sum_{\substack{i,k=1 \\ i \neq k}}^{4} (m_i + m_k)^2 - 2\sum_{i=1}^{4} m_i^2$$

Such a choice of point $s_0$ allows one to replace the integral Eqs. (3) - (8) (Fig. 1) by the algebraic equations (19) - (24) respectively:

$$\alpha_1 = \lambda_1 + 3\widehat{J}_2(3,1,1)\alpha_5 + 3\widehat{J}_2(3,1,2)\alpha_6 + 2\widehat{J}_2(3,1,3)\alpha_4 + 2\widehat{J}_1(3,3)\alpha_1 + 2\widehat{J}_1(3,2)\alpha_3 + \\ + 2\widehat{J}_1(1,1)\alpha_1 + 2\widehat{J}_1(1,2)\alpha_2 \tag{19}$$

$$\alpha_2 = \lambda_2 + 6\widehat{J}_2(3,2,1)\alpha_5 + 2\widehat{J}_2(3,2,3)\alpha_4 + 2\widehat{J}_1(3,3)\alpha_2 + 2\widehat{J}_1(3,1)\alpha_3 + 4\widehat{J}_1(2,2)\alpha_2, \tag{20}$$

$$\alpha_3 = \lambda_3 + 6\widehat{J}_2(1,2,1)\alpha_5 + 6\widehat{J}_2(1,2,2)\alpha_6 + 4J_1(2,3)\alpha_1 + 4\widehat{J}_1(1,3)\alpha_2 \tag{21}$$

$$\alpha_4 = \lambda_4 + 8\widehat{J}_3(3,3,2)\alpha_2 + 4\widehat{J}_3(3,1,3)\alpha_1, \tag{22}$$

$$\alpha_5 = \lambda_5 + 4\widehat{J}_3(1,1,2)\alpha_3 + 2\widehat{J}_3(1,2,3)\alpha_2 + 2\widehat{J}_3(1,1,3)\alpha_1, \tag{23}$$

$$\alpha_6 = \lambda_6 + 4\widehat{J}_3(2,2,1)\alpha_3 + 2\widehat{J}_3(2,1,3)\alpha_1 + 2\widehat{J}_3(2,2,3)\alpha_2 \tag{24}$$

We use the functions $J_1(l, p)$, $J_2(l, p, r)$, $J_3(l, p, r)$ ($l, p, r = 1, 2, 3$):

$$J_1(l, p) = \frac{G_l^2(s_0^{12}) B_p(s_0^{13})}{B_l(s_0^{12})} \int_{(m_1+m_2)^2}^{\Lambda_l} \frac{ds'_{12}}{\pi} \frac{\rho_l(s'_{12})}{s'_{12} - s_0^{12}} \int_{-1}^{+1} \frac{dz_1}{2} \frac{1}{1 - B_p(s'_{13})}, \tag{25}$$

$$J_2(l,p,r) = \frac{G_l^2(s_0^{12})G_p^2(s_0^{34})B_r(s_0^{13})}{B_l(s_0^{12})B_p(s_0^{34})} \times$$

$$\times \int_{(m_1+m_2)^2}^{\Lambda_l} \frac{ds'_{12}}{\pi} \frac{\rho_l(s'_{12})}{s'_{12}-s_0^{12}} \int_{(m_3+m_4)^2}^{\Lambda_p} \frac{ds'_{34}}{\pi} \frac{\rho_p(s'_{34})}{s'_{34}-s_0^{34}} \int_{-1}^{+1} \frac{dz_3}{2} \int_{-1}^{+1} \frac{dz_4}{2} \frac{1}{1-B_r(s'_{13})} , \quad (26)$$

$$J_3(l,p,r) = \frac{G_l^2(s_0^{12},\tilde{\Lambda})B_p(s_0^{13})B_r(s_0^{24})}{1-B_l(s_0^{12},\tilde{\Lambda})} \frac{1-B_l(s_0^{12})}{B_l(s_0^{12})} \frac{1}{4\pi} \times$$

$$\times \int_{(m_1+m_2)^2}^{\tilde{\Lambda}} \frac{ds'_{12}}{\pi} \frac{\rho_l(s'_{12})}{s'_{12}-s_0^{12}} \int_{-1}^{+1} \frac{dz_1}{2} \int_{-1}^{+1} dz \int_{z_2^-}^{z_2^+} dz_2 \frac{1}{\sqrt{1-z^2-z_1^2-z_2^2+2zz_1z_2}} \frac{1}{[1-B_p(s'_{13})][1-B_r(s'_{24})]} \quad (27)$$

The other choices of point $s_0$ do not change essentially the contributions of $\alpha_1$, $\alpha_2$, $\alpha_3$, $\alpha_4$, $\alpha_5$ and $\alpha_6$; therefore we omit the indexes $s_0^{ik}$. Since the vertex functions depend only slightly on energy it is possible to treat them as constants in our approximation. The integration contours of function $J_1, J_2, J_3$ are given in paper [25].

The solutions of the system of equations are considered as:

$$\alpha_i(s) = F_i(s,\lambda_i)/D(s), \quad (28)$$

where zeros of $D(s)$ determinants define the masses of bound states of pentaquark baryons. $F_i(s,\lambda_i)$ are the functions of $s$ and $\lambda_i$. The functions $F_i(s,\lambda_i)$ determine the contributions of subamplitudes to the pentaquark baryon amplitude.

The poles of the reduced amplitudes $\alpha_1$, $\alpha_2$, $\alpha_3$, $\alpha_4$, $\alpha_5$, $\alpha_6$ correspond to the bound states and determine the masses of $P_{sss}^0$ ($sssu\bar{d}$) pentaquarks (Fig. 1). The model in considerations, has only one new parameter as compared to our previous paper (Ref. 15). The dimensionless parameter is shift of mass $\Delta = 100$ MeV. It is determined by fixing of $P_{sss}^0$ pentaquark mass (2250 MeV). The cutoff parameters coincide with those in Ref. [15]: $\Lambda_{0^+} = 16.5$ and $\Lambda_{1^+} = 20.12$ for the diquarks with $0^+$ and $1^+$, respectively. The cutoff parameter for the mesons is equal to $\Lambda_{0^+} = 16.5$. The dimensionless parameter gluon coupling constant is similar to the paper [15]. The calculated mass values of low-lying pentaquarks $P_{sss}^0$ are shown in Table I. The masses of $P_{sss}^0$ pentaquarks with the quantum numbers isospin $I = 0$ and spin-parity $J^P = \frac{1^{\pm}}{2}, \frac{3^{\pm}}{2}$ are predicted. The mass of $P^0_{sss}$ pentaquarks with positive parity is smaller than the mass of pentaquarks with negative parity. This depends on the different interaction in the diquark channels. It should be noted, that the calculated masses of low-lying strange pentaquarks agree with the data [17] (Table I).

The interesting research is the consideration of the other states $qqqQ\bar{Q}$ with $q$ quarks and $Q$ heavy quark ($Q = c, b$).

Researchers have made great progress using lattice gauge theory, which allows a numerical treatment of QCD, to determine the spectroscopy of strongly interacting particles [26]. But, theory does not provide us with a good understanding of the heavy multiquark particles.

Table I. Strange ($S = -3$) pentaquark masses (MeV)

| $J^P$ | Mass, MeV |
|---|---|
| $\frac{1}{2}^+$ | 2250 (2250) |
| $\frac{3}{2}^+$ | 2490 (2470) |
| $\frac{1}{2}^-$ | 2344 (2380) |
| $\frac{3}{2}^-$ | 2634 ( – ) |

Parameters of model: quark mass $m_{u,d}$ = 410 MeV, $m_s$ = 557 MeV;
cut-off parameters $\Lambda_{0^+}$ =16.5, $\Lambda_{1^+}$ =20.12; gluon coupling constant $g$ =0.456, $\Delta$ =100 MeV.

Figure captions

Fig.1. Graphic representation of the equations for the five-quark subamplitudes $A_k$ ($k$ =1-6) corresponds to the strange pentaquarks ($S = -3$).

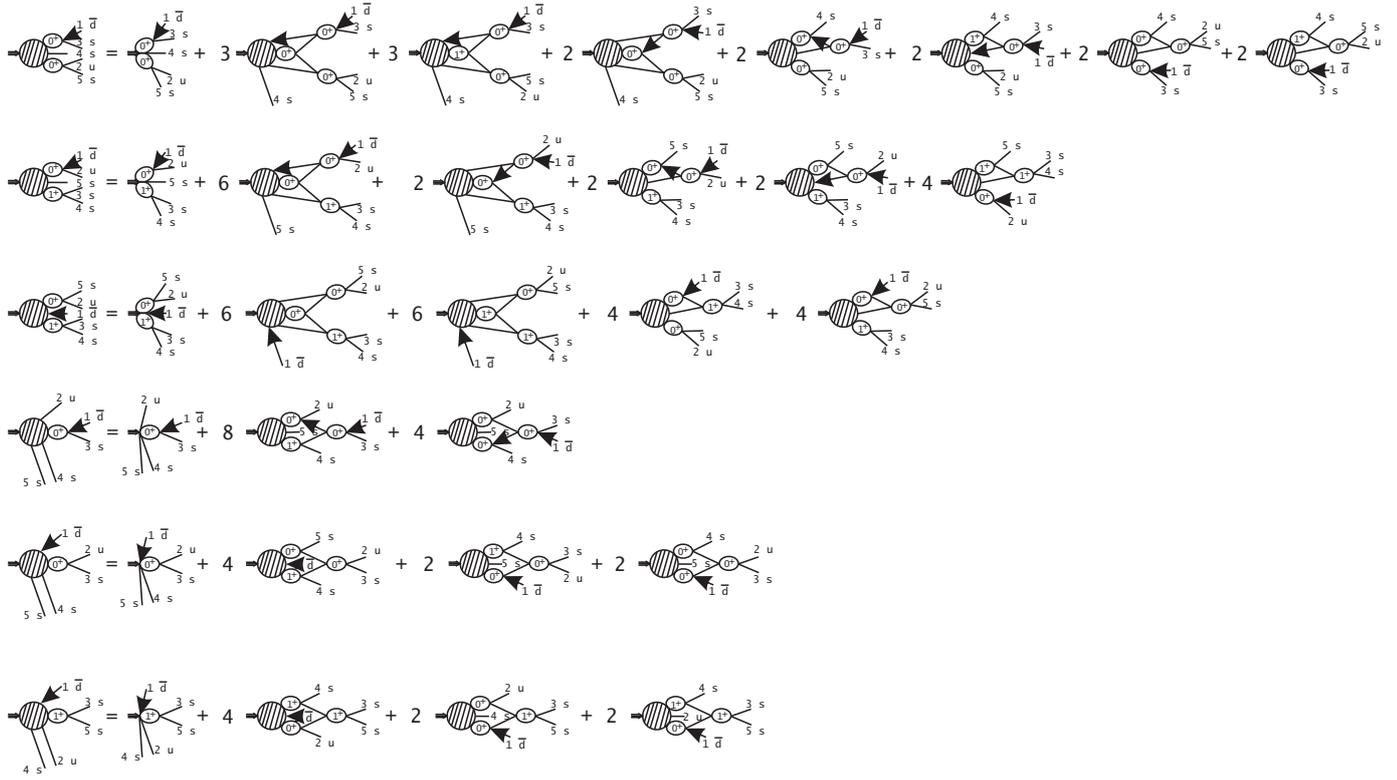

Fig.1